\documentclass[]{article}
\usepackage{graphicx}
\usepackage[latin1]{inputenc}
\usepackage{amsmath}
\usepackage{comment}
\usepackage{authblk}   

\usepackage{xcolor}
\usepackage[urlcolor=blue]{hyperref}      
\hypersetup{
    colorlinks = true,                    
    citecolor = {blue},
    linkcolor = {purple},
           }

\title{\bf When to end a lock down?  How fast must vaccination campaigns 
proceed in order to keep health costs in check?}

\author[$\dagger$]{Claudius Gros}
\author[$\star$]{Thomas Czypionka}
\author[$\ddag$]{Daniel Gros}

\affil[$\dagger$]{Institute for Theoretical Physics, 
                  Goethe University Frankfurt, Germany}
\affil[$\star$]{Institute for Advanced Studies, Vienna; 
                London School of Economics and Political Science, London}
\affil[$\ddag$]{CEPS (Centre for European Policy Studies), Brussels, Belgium} 

\begin{document}

\maketitle
\begin{abstract}
We propose a simple rule of thumb for countries
which have embarked on a vaccination campaign while 
still facing the need to keep non-pharmaceutical 
Interventions (NPI) in place because of the ongoing 
spread of SARS-CoV-2. If the aim is to keep the death 
rate from increasing, NPIs can be loosened when it 
is possible to vaccinate more than twice the growth 
rate of new cases. If the aim is to keep the pressure 
on hospitals under control, the vaccination rate 
has to be about 4 times higher.
These simple rules can be derived 
from the observation that the risk of death or a severe
course requiring hospitalisation from a 
Covid-19 infection increases exponentially with age 
and that the sizes of 
age cohorts decrease linearly at the top of
the population pyramid. 
Protecting the over 60-year-olds, which constitute approximately one quarter of the population in 
Europe (and most OECD countries), reduces the 
potential loss of life by 95 percent.

\end{abstract}


\newpage
\section{Introduction}

Infection with the SARS-CoV-2 virus represents a 
serious health risk. A number of studies have by now established that this risk increases exponentially with age (see \cite{levin2020assessing} for a survey and metastudy). This applies both to the risk of dying and 
to the risk of a severe course of the illness 
requiring hospitalisation \cite{CDC_age_risk}. 

Vaccination 
campaigns have therefore prioritized the elderly 
(along with health workers). Countries with active 
outbreaks have in general introduced NPIs to restrict 
mobility and reduce the number of personal contacts.\cite{chen2020tracking} 
However, NPIs have a high social and economic cost. 
This cost arises both through the limitations to 
economic activity \cite{cutler2020covid} and the 
health costs from those catching the virus and 
becoming infected \cite{gros2020containment}. 
It is 
thus key to understand when NPIs can be lifted safely. 
This has lead to a race between mass scale vaccination campaigns 
and the disease, which continued to spread and evolve.  
\cite{dye2020scale} provides an analysis of the scale 
and the spread of the disease in 2020.  \cite{zheng2021mathematical} provides a mathematical model. 
\cite{gnanvi2021reliability} provides an overview 
of the statistical models used to forecast the 
spread of the disease.  A notable application of statistical modelling to the European experience is provided by \cite{agosto2021monitoring}  and (using 
a Poisson process) \cite{agosto2020poisson}.  
Moreover, new variants have appeared \cite{bauer2021relaxing}, with new and potentially 
more infectious strands \cite{davies2021estimated}.

The ultimate aim of vaccination campaigns is to 
achieve `herd' or population immunity, i.e.\ a 
state in which the share of the population that 
has been immunized by infection or vaccination 
brings down the effective reproduction number 
$R\textsubscript{eff}$ below unity without further 
NPIs needed \cite{linka2020reproduction}. However, 
vaccination takes time, mostly due to the difficulties 
in ramping up the production of vaccines \cite{gros2021incentives},
but also in part also due to bottlenecks 
in distribution and implementation. 
A core issue for policy makers therefore is 
the point during the vaccination campaign 
at which NPIs can be lifted. 

We provide a general rule to answer this question.  
Our proposed rule is based only on observable data 
and is the result of combining two simple relationships
\begin{itemize}
\item
The age-dependency of the case fatality and the 
hospitalization rate, which has been established 
to increase approximately
exponentially with age \cite{levin2020assessing}.
\item
The population structure for the elderly, which is 
to first order linear at the top, which means that 
the size of age cohorts increases gradually top-down 
from the maximum age (about 100 years)
\cite{grundy2017population}.
\end{itemize}

We concentrate on observable outcomes like death or 
hospitalisation because their number constitutes 
a key determinant for the imposition of lock-downs 
\cite{ferraresi2020great} and other NPIs, which 
come with severe economic and social costs in terms 
of lost output and employment \cite{dreger2021lockdowns} 
as well as indirect impacts on health and well-being \cite{Brodeur}.

The combination of these two elements leads 
to a simple rule 
of thumb to recognise the 'sweet point' at 
which NPIs can 
be loosened while still 'flattening the curve'.
We then add a third element, namely 
\begin{itemize}
\item
The functional dependency of daily vaccinations rates,
which are observed to increase in most countries 
approximately linearly over time. 
\end{itemize}
This third element allows one to predict 
the stage in the vaccination process at 
which the 'sweet point' will be reached.
We concentrate throughout mostly on the 
case of Europe, because this is the region 
in which a resurgence of infections coincided 
with an, initially at least, sluggish vaccination campaign.

\section{Modeling framework}


\begin{figure}[!t]
\centering
\includegraphics[width=0.95\columnwidth]{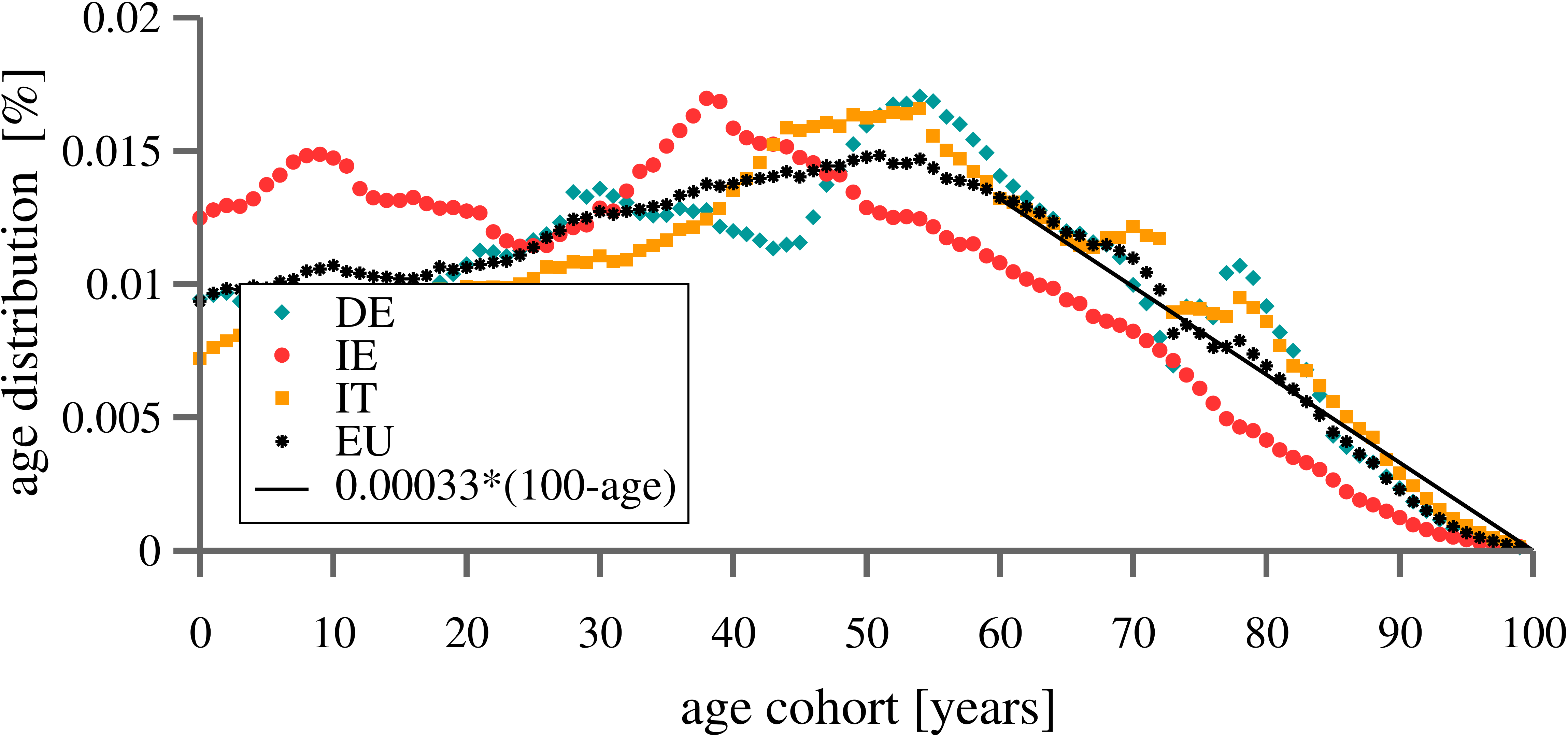}
\newline
\includegraphics[width=0.95\columnwidth]{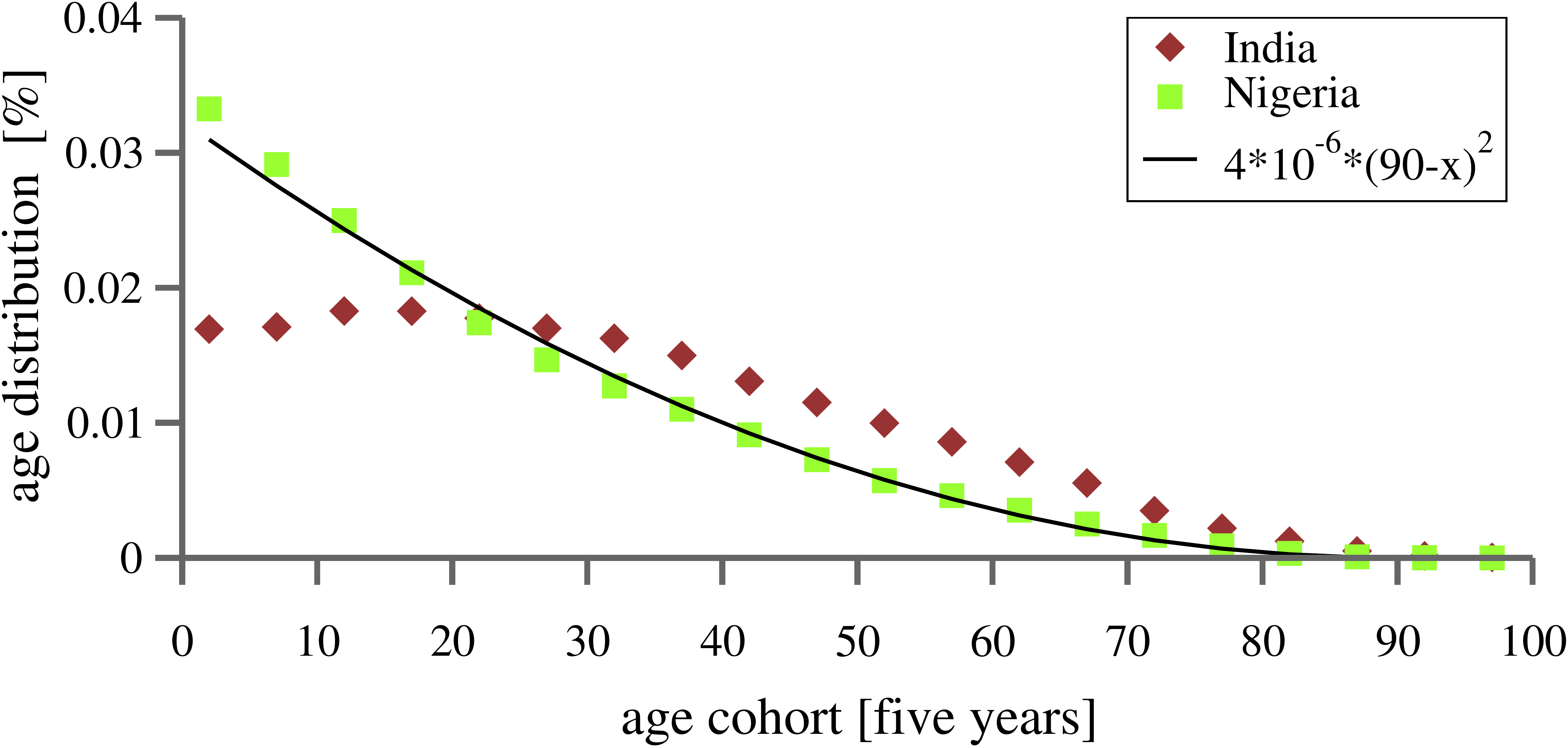}
\caption{{\bf Age distributions.}
(Top) Fractional yearly age cohorts for 
Germany (DE), Ireland (IE), Italy (IT) and 
EU27 (EU). Also shown is a linear interpolation
for the EU data (black line), for ages 60 and 
above. Data from \cite{eurostat21}.
(Bottom) Five-year age cohorts for India
and Nigeria. A quadratic interpolation to
the data for Nigeria is also provided.
}
\label{fig_ageDistribution}
\end{figure}

\subsection{Age-dependency of the health risks of Covid-19 infections}

We concentrate on two observable health risks: 
death and hospitalisation

It has been widely documented that the risk to 
die from a Covid-19 infection rises strongly 
with age. A meta-study suggests an exponential 
relationship \cite{levin2020assessing}, which
can be parameterized as
\begin{equation}
\mathrm{IFR}\approx 0.01\mbox{e}^{-7.529+0.121*a}\,
\sim \mbox{e}^{a/a_0}
\qquad\quad
a_0 = 8.26
\label{IFR}
\end{equation}
where $a\in[0,100]$ is the age cohort. The infection 
fatality rate $\mathrm{IFR}\in[0,1]$ is very high
for $a=100$, namely $\mathrm{IFR}(100)=0.93$.

The constant $a_0$ denotes the half life age difference 
in terms of mortality. To be more precise, for an age 
difference of $8.26$ years the risk increases by a 
factor of $e=2.78$. The risk doubles for an 
age difference of 5.7 years.\footnote{Statistics 
from the CDC of the US show that the over 85 years 
old have a Covid-19 mortality rate 7,900 times higher 
than that of the age group 5-17. This translates 
also into a doubling age difference of about 8 
years (= (85-11)/ln(7900)\,\cite{CDC_age_risk}.}

Limiting the strain on health systems has been another constant concern of policy makers.  The risks of requiring hospitalisation is also age specific. Statistics 
from the US CDC show that people aged 85 years 
and older face a risk of hospitalisation from 
Covid-19 infections about 95 times higher 
than that of the age group 5-17. This translates 
also into a doubling age difference of about 16 
years (= (85-11)/ln(95), which we denote as $a_{0h}$ about double of that of the fatality risk. 

In the following we will concentrate on decision rules designed to limit fatalities, but we will also refer to consequence of this somewhat different age dependency when the aim is to limit the pressure on the health system;

\subsection{Age pyramid}

We present the age pyramid for a range of selected countries
in Fig.~\ref{fig_ageDistribution},
where the age cohorts are given as a share of the 
entire population. One notices that the age pyramid 
closes generically quadratically at the very top. For
EU countries,he relevant range for the quadratic 
dependency is however restricted, applying only to 
ages around 85 and above.

Most Covid-19 vaccination campaigns started by
following top-down strategies to varying degrees  
\cite{castro2021prioritizing}. For the time being we concentrate on EU countries, for which
the respective age pyramids can be approximated
linearly, as shown in Fig.~\ref{fig_ageDistribution}. 
This approximation is intended as an overall fit 
to ages 60 and above. The case of countries with
younger populations, like India and Nigeria, will
be treated later. \cite{ritchie2019age}

In the following we set the maximum age to zero,
counting down from 100. The actual age is then
$100-x$. The age density, denoted by $\rho(x)$,
varies by cohorts. Countries with population 
pyramids closing linearly at the top are 
described by
\begin{equation}
\rho(x) = px, 
\qquad\quad
\int_0^{\delta\!A} \rho(x)\,dx = v,
\qquad\quad
v = \frac{p(\delta\!A)^2}{2}\,,
\label{rho_x}
\end{equation}
where $v$ is the number of people (relative to
the total population) vaccinated top-down to 
an age difference $\delta\!A$.

Fitting to aggregate European demographic data yields
$p\approx0.00033$, as shown in 
Fig.~\ref{fig_ageDistribution}, with little 
difference across the major EU member countries.
This value for $p$ refers to the case that 
$\rho(x)$ is measured relative to the total 
population, in terms of percentiles of cohorts by year. For example, with this value of $\rho(x)$ the share of the 70-year-olds in the total population is equal to about 0.01 $=0.00033*30$. 

The linear approximation holds well down to 
60 years. Below this age, the size of the cohort 
no longer increases (and even falls in some countries, 
like Italy). Here we concentrate on the age cohorts 
from 60 years up, which are the ones subject to the 
highest mortality risk, constituting the largest 
proportion of the overall loss of life. The  case 
of Germany illustrates this proposition. Taking 
into account the combined effect of (\ref{IFR}) 
and the age distribution, as presented in 
Fig.~\ref{fig_ageDistribution}, one finds that 
about 1.5 million people above the age of sixty die in 
the hypothetical scenario that the entire population 
would be eventually infected with 
SARS-CoV-2. By contrast, the 
fatality count would include only 75 thousand 
below sixty, i.e. by a factor of twenty. We thus feel 
justified concentrating our analysis on the age 
cohorts above 60, for which the population pyramid 
is approximately linear. 

People over 60 years of age account
for about 26 percent of the total population of 
the EU, with their shares ranging from 20 percent 
in the case of Ireland to 29 percent in the case 
of Italy. This implies that vaccinating about one 
fourth of the population will avoid 95 
percent of fatalities ($19/20)$.
This calculation based solely on age obviously 
represents an approximation.
Due to vaccine hesitancy, non-responders 
and people with contraindications to 
the vaccine, the uptake among the 
elderly could be less than 100 percent. But 
these factors are also present among all age 
groups, reducing thus the overall benefit from 
a vaccination campaign, but not necessarily 
the advantage of age-sensitive vaccination.
Vaccine hesitancy is in particular likely to be 
lower among the elderly, implying that the 
share of the benefits from offering vaccination 
to the elderly first might be even higher than 
the 95\% suggested on demographic 
considerations alone. A factor suggesting
otherwise may however be `long Covid'
\cite{altmann2021decoding}.

There is also evidence that immunity wanes more 
quickly at higher ages \cite{hansen2021assessment}, 
implying that the re-infection risk is higher 
for the elderly. This effect, however, plays out on a time scale beyond that of most current vaccination campaigns.

\begin{figure}[!t]
\centering
\includegraphics[width=0.95\columnwidth]{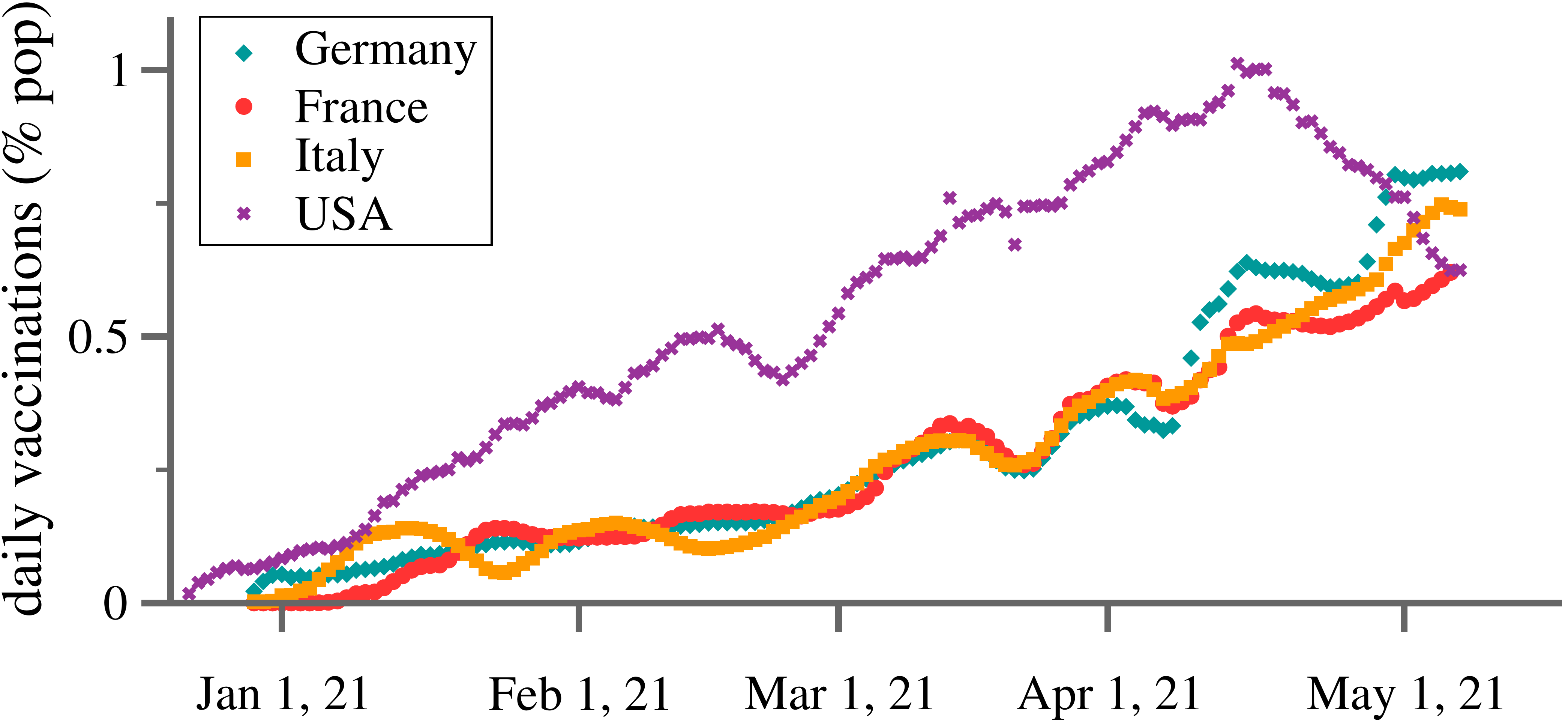}
\caption{{\bf Growth of daily vaccination rates.}
Per million daily vaccinations in Italy, France, Germany
and the United States (USA).
The growth is roughly linear, modulo in part substantial
fluctuations. Data smoothed over seven days, from \cite{OWID_covid19}. 
}
\label{fig_dailyVaccinations}
\end{figure}

\section{Flattening the health cost curve}

Putting the two basic elements
-- the exponential age-dependency of the case fatality rate,
the linear functionality of population pyramid, --
together, we proceed to calculate the 'sweet spot' at 
which the vaccination campaign can stabilise health costs.

We concentrate on the \emph{growth} of fatalities 
because this is the key concern for policy makers. 
But we also show that our approach can easily 
be used to address the generic concern  'flatten 
the curve', i.e.\ 
to prevent an explosive increase in hospitalizations 
which could overwhelm health systems. 

Throughout this section we take the growth of infections as given.  In section \ref{Vaccination hesitancy} we discuss how the results are modified by the feedback between vaccination and infections. 

\subsection{Vaccination and infection fatality rate}

We assume that full vaccination provides 
a high level of protection against severe 
illness and death, as confirmed not only by  
trial data \cite{mahase2020covid}, but also 
by real world 
application \cite{aran2021estimating,dagan2021bnt162b2}. 
For our model we furthermore assume that vaccination 
is allocated strictly by age, starting with 
the oldest. In practice the situation is
more complicated. Firstly, because a substantial
fraction of the available vaccine is reserved
for potential spreaders in most countries
\cite{castro2021prioritizing}, independent of
their age. Secondly, one needs to distinguish
between people having received one or two shots.
Both effects could be incorporated into the
framework developed here. In order to clarify
the mechanisms at work, we study in the
following the idealized situation that
`vaccinated' implies full protection.

\section{Medical balancing condition}

In a first step we derive a medical balancing 
condition that is generically valid, viz for
all types of population pyramids. In a second
step we will make use of the fact that population
pyramids may often be well approximated linearly
at the top. As before, we do not work with the nominal 
age $a\in[0,100]$, but with the relative age $x$, 
as measured top-down, $a=100-x$, denoting by 
$\rho(x)$ the respective population density. The
fraction $v$ of people vaccinated top-down
to an age difference $\delta\!A$ is then
\begin{equation}
v = \int_0^{\delta\!A} \rho(x)dx,
\qquad\quad
\frac{dv}{dt} = \rho(\delta\!A)\frac{d(\delta\!A)}{dt}\,,
\label{v_delta_A_dt}
\end{equation}
where included also a differential relation
with regard to the vaccination speed $\dot v$.
Above expression reduces to 
$v=p(\delta\!A)^2/2$ for a population pyramid
closing linearly at the top, viz
for $\rho(x)=px$. 

A key variable for policy makers is the medical load, i.e. the number of patients requiring hospitalisation 
and maybe even intensive care or who die. 
We start by considering the mortality risk. The 
medical load or cost for society at any given 
state of the vaccination
campaign can then be expressed as as the number 
of people who risk dying each time period:
\begin{equation}
C^{\rm med} =  \left[\int_{\delta\!A}^\infty 
\rho(x)\,\mbox{e}^{-x/a_0}dx\right]I(t)\,,
\label{C_med_general}
\end{equation}
where it is was assumed that fully vaccinated people 
will not get seriously ill anymore.
Daily Covid-19 fatalities do not increase when
$C^{\rm med}$ remains stable, viz when the
balancing condition $dC^{\rm med}/dt=0$ holds.
Setting the time derivative of the right-hand 
side of (\ref{C_med_general}) to zero leads
to the condition that:
\begin{equation}
\frac{d\delta\!A}{dt}\left[
\rho(\delta\!A)\,\mbox{e}^{-\delta\!A/a_0}
\right] = \left[\int_{\delta\!A}^\infty 
\rho(x)\,\mbox{e}^{-x/a_0}dx\right]
\frac{\dot I}{I}\,.
\label{balancing_condition_general_00}
\end{equation}
With the expression for $\dot v$ given in 
(\ref{v_delta_A_dt}), the left-hand side of
(\ref{balancing_condition_general_00})
reduces to $\dot{v}\exp(-\delta\!A/a_0)$.
We then have
\begin{equation}
\dot{v} = \mbox{e}^{\delta\!A/a_0} 
\left[\int_{\delta\!A}^\infty 
\rho(x)\,\mbox{e}^{-x/a_0}dx\right]
\frac{\dot I}{I}\,.
\label{balancing_condition_general}
\end{equation}
This expression is valid for all
population densities $\rho(x)$, in
particular also for the tabulated
population density of a given country.

\subsection{Linear population pyramid}

Evaluating the integral in
(\ref{balancing_condition_general})
for the linear case, when $\rho(x)=px$,
one has
\begin{eqnarray}
\nonumber
p\int_{\delta\!A}^\infty x\,\mbox{e}^{-x/a_0}dx 
&=&  -(pa_0^2) \left[1+\frac{x}{a_0}\right]
\mbox{e}^{-x/a_0}\Big|_{\delta\!A}^\infty
\\ &=&  (pa_0^2) \left[1+\frac{\delta\!A}{a_0}\right]
\mbox{e}^{-\delta\!A/a_0}\,.
\label{integral_linear}
\end{eqnarray}
With
\begin{equation}
v_0 = \frac{a_0^2p}{2},
\qquad\quad
v = \frac{p\delta\!A^2}{2},
\qquad\quad
\frac{\delta\!A}{a_0} = \sqrt{\frac{v}{v_0}}
\label{relations}
\end{equation}
one then finds that
(\ref{balancing_condition_general})
reduces to 
\begin{equation}
\frac{\dot{v}}{v_0} = 2
\left(1+\sqrt{\frac{v}{v_0}}\,\right)\frac{\dot{I}}{I}\,,
\label{balancing_linear}
\end{equation}
where $\dot{I}/I$ is the (relative) increase
of the incidence.

\section{Consequences of the balancing condition}

\subsection{The sweet spot for an old country}

When the population pyramid is linear at
the top (as it is typically for ageing societies), 
the balancing condition (\ref{balancing_linear}) 
leads to a simple rule of thumb, given that $v_0$ 
is equal to about 0.01, or one percent:
\begin{quote}
{\em``For every proportional increase $\dot I/I$
of the incidence, one needs to vaccinate an 
additional percentage of \emph{at least} 
twice that amount in order to outrun the virus.''}
\end{quote}
This lower bound holds for $v\to0$, becoming
larger when vaccination progresses. 
The balancing condition thus provides a simple 
decision rule based on observable data (infections 
and vaccinations). Once this 'sweet spot' has been 
reached, NPIs can be gradually loosened without risking 
an increase in fatalities.

\subsection{The sweet spot: hospitalizations}

The balancing conditions derived above describe 
the point at which the number of fatalities 
(per unit of time) stops increasing. One key 
parameter used to derive these results was $a_0$,
which denotes the half live age difference in 
terms of mortality. For the risk of hospitalization 
this parameter, $a_{0h}$, about twice as high, 
but the functional form remains the same.   

For a linear age pyramid this implies that the 
evolution of hospitalisations can be described 
by the same balancing condition,
\begin{equation}
\frac{\dot{v}}{v_{0h}} = 2
\left(1+\sqrt{\frac{v}{v_{0h}}}\,\right)\frac{\dot{I}}{I}\,,
\label{balancing_linear_h}
\end{equation}
where the subscript $_h$ indicates that the value 
is calculated with $a_{0h}$ instead of $a_0$.  
This changes two of the three relationships 
in equation (\ref{relations}):
\begin{equation}
v_{0h} = \frac{a_{0h}^2p}{2},
\qquad\quad
v = \frac{p\delta\!A^2}{2},
\qquad\quad
\frac{\delta\!A}{a_{0h}} = \sqrt{\frac{v}{v_{0h}}}
\label{relations_h}
\end{equation}
The (relative) increase of the incidence, 
$\dot{I}/I$ and $v$ are not affected. 

Comparing (\ref{relations_h}) and (\ref{relations}) 
shows that $v_{0h}/v_0$ is equal to $a_{0h}/a_0$.  
It was documented above that the mean doubling age 
for hospitalizations is about two times larger than 
that for deaths (16 instead of 8 years). This 
implies that $v_{0h}$ is about four times larger 
than $v_0$. It follows that at low vaccination 
rates ($v$ small) the balancing condition is four 
times more stringent if the aim is to keep 
hospitalizations from increasing than if the 
aim is to keep fatalities in check, i.e.\ 
the decision rule would become:

\begin{quote}
{\em``For every proportional increase $\dot I/I$
of the incidence, one needs to vaccinate an 
additional percentage of \emph{at least} 
eight times that amount in order to keep the 
pressure on hospitals constant.''}
\end{quote}

A policy that takes into account the need to limit 
the pressure on health systems would thus loosen 
NPIs much later than a policy which concentrates only
on fatalities. The difference is a factor of four at 
low vaccination rates, but it diminishes as a higher 
proportion is vaccinated (as $v$ increases) because the
higher value of $a_{0h}$ enters the denominator of 
the second term in the brackets on the right hand side 
of equation (\ref{balancing_linear_h}).

%
%

\subsection{The sweet spot for a young country}

As documented below, younger countries have a 
top of the population pyramid which closes 
quadratically. In this case the size of the 
cohorts counting from age 100 down can be described by:
\begin{equation}
\rho(x)=qx^2,
\quad\quad
v = \frac{q\delta\!A^3}{3},
\quad\quad
\frac{\delta\!A}{a_0} = \left(\frac{3v}{a_0^3q}\right)^{1/3},
\quad\quad
v_1 \equiv \frac{a_0^3q}{3}
\label{rho_xx}
\end{equation}
and
\begin{equation}
q\int_{\delta\!A}^\infty x^2\,\mbox{e}^{-x/a_0}dx 
=  -(qa_0^3) 
\left[2+\frac{2x}{a_0}+\frac{x^2}{a_0^2}\right]
\mbox{e}^{-x/a_0}\Big|_{\delta\!A}^\infty,
\label{integral_quadratic}
\end{equation}
one obtains the balancing condition
\begin{equation}
\frac{\dot v}{v_1} = 3\left[
2 + 2\left(\frac{v}{v_1}\right)^{1/3}
+ \left(\frac{v}{v_1}\right)^{2/3}
\right]\frac{\dot I}{I}
\label{balancing_quadratic}
\end{equation}
A fit to the population pyramid of Nigeria
yields $q=4\cdot10^{-6}$, viz $v_1=a_0^3q/3=0.0002$,
see (\ref{IFR}) and Fig.~\ref{fig_ageDistribution}.
The reference fraction of vaccinated is hence
exceedingly small, $v_1\sim 0.02\%$

A generalization
to the case of a quadratic population pyramid,
as given by (\ref{balancing_quadratic}), yields
a factor $6v_1/v_0\approx 0.12$ in the limit
$v\to0$, instead of two (given that $v_0$ corresponds
to about one percent). At the start, vaccination 
campaigns have hence strong control capabilities in
countries with young populations, viz when the
population pyramid is quadratic at the top. This
advantage decreases rapidly, however, with 
the progress of vaccination. To be concrete, dividing
equation (\ref{balancing_linear}) by
(\ref{balancing_quadratic}) one obtains a factor
of $1.7$ at $v=0.1$, when ten percent of the
population has been vaccinated top-down.

\subsection{Connection with the evolution of the pandemic}

The rate of increase in the number of infected will 
be affected by the fraction of the population already 
vaccinated. 
The evidence regarding the impact of vaccines 
on the spread of infections is less clear than the impact of vaccination on the risk of death or a severe course 
\cite{lipsitch2021interpreting}. Some results point to a reduction in transmission of 40 percent \cite{harris2021effect} whereas others report a much higher impact among the vulnerable population in long-term care facilities  \cite{de2021high}.
Any impact of vaccination on infectiousness 
would not change the balancing condition 
(\ref{balancing_linear}), which applies in 
general. Moreover, with vaccination reducing 
infectiousness, whatever the size of the impact, the growth of the disease spread slows down, ceteris 
paribus, (to a lower value for $\dot I/I$), 
making it easier to reach the point where the curve `flattens'. 

The relative increase $\dot I/I$ appearing
on the right-hand side of (\ref{balancing_linear})
could be evaluated using a dynamic epidemiological
model. For the SIR model \cite{gros2015complex},
the simplest case, we have
$\dot I/I = gS-\lambda$, where $g$ and $\lambda$ are
the reproduction and the recovery rates, respectively.
The reservoir of susceptibles $S$ decreases with the 
fraction of the population that is immunized, either
because of having been infected, or due to being
vaccinated. We leave this approach for further
studies, concentrating here on data-driven considerations.

The dynamics of the growth of a pandemic is 
studied in a large body of literature. One of 
the first to employ the 
canonical model was \cite{ferguson2020imperial}.  \cite{agosto2021monitoring} employ statistical 
models to predict the contagion curve and the associated 
reproduction rate using a Poisson process. See also
\cite{harvey2020time} for time series models and
\cite{shinde2020forecasting} for a survey.

We thus conclude that the position of the 
'sweet spot' in terms of observed infections is not
changed if one takes into account the impact of 
vaccinations on the spread of the disease. With 
continuing vaccinations the sweet spot becomes 
a turning point because once it has been 
reached, vaccination rates  could plateau, 
but the condition for deaths to fall would 
remain fulfilled, as additional vaccinations 
reduce the growth of infection, ${\dot I}/{I}$, 
which is the
determining factor on the right-hand side of 
(\ref{balancing_linear}).
 

\section{Vaccination hesitancy}
\label{Vaccination hesitancy}

In deriving (\ref{balancing_condition_general}),
we assumed full vaccination of everybody aged
above $a=100-\delta\!A$. In practice, there will
be a certain hesitancy $h(x)\in[0,1]$, meaning
that a fraction $h(x)$ remains non-vaccinated, 
even after vaccines have become available to the 
age cohort $x$. The age specific indicator $h=h(x)$
is used here as a general term, including people 
that are not willing to be vaccinated, together
with the fraction of the population that cannot
be vaccinated due to medical contra-indications. 
People not developing an immune response despite 
being fully vaccinated are also subsumed under 
$h(x)$.
  
In Europe, the fraction $h(x)$ of not vaccinated
starts close to zero in most countries for the 
very old, remaining in general below 20 per cent 
down to age 60.\footnote{\href{https://www.ecdc.europa.eu/en/publications-data/data-covid-19-vaccination-eu-eea}{https://www.ecdc.europa.eu/en/publications-data/data-covid-19-vaccination-eu-eea}}
It is straightforward to generalize the derivation of
the medical balancing condition to the case $h(x)>0$. 
For this the term:
\begin{equation}
c_0I(t)\int_0^{\delta\!A} h(x)\rho(x)\mbox{e}^{-x/a_0}dx
\label{C_Med_term_hesitancy}
\end{equation}
needs to be added to the medical cost $C^{\rm med}$,
see (\ref{C_med_general}). One finds
\begin{equation}
\dot{v}\big(1-h(\delta\!A)\big) = \mbox{e}^{\delta\!A/a_0} 
\left[
\int_{\delta\!A}^\infty \rho(x)\,\mbox{e}^{-x/a_0}dx +
\int_0^{\delta\!A} h(x)\rho(x)\,\mbox{e}^{-x/a_0}dx
	\right]
\frac{\dot I}{I}\,,
\label{balancing_condition_hesitancy}
\end{equation}
which reduces to (\ref{balancing_condition_general})
when $h(x)\to0$. Clearly, it becomes more difficult
to retain medical balancing when vaccine hesitancy
is large, which is however not the case for the elderly.

For the case of age-independent hesitancy, $h(x)\equiv h$,
one can evaluate both the first and the second
integral on the right-hand side of
(\ref{balancing_condition_hesitancy}). The latter is
\begin{eqnarray}
\nonumber
hp\int_0^{\delta\!A} x\,\mbox{e}^{-x/a_0}dx 
&=&  -h(pa_0^2) \left[1+\frac{x}{a_0}\right]
\mbox{e}^{-x/a_0}\Big|_0^{\delta\!A}
\\ &=& - h(pa_0^2) \left[1+\frac{\delta\!A}{a_0}\right]
\mbox{e}^{-\delta\!A/a_0} + h(p a_0)^2\,.
\label{integral_linear_hesitancy}
\end{eqnarray}
The first term on the right-hand side is $(-h)$
times the obtained for the case without hesitancy.
All together we hence have
\begin{equation}
\frac{\dot{v}}{v_0} = 
2 \left(1+\sqrt{\frac{v}{v_0}}\,\right)\frac{\dot{I}}{I} 
+
2h\frac{\mbox{e}^{\delta\!A/a_0}}{1-h}\,\frac{\dot{I}}{I}
=  2\left[1+\sqrt{\frac{v}{v_0}}+
h\frac{\mbox{e}^{\delta\!A/a_0}}{1-h}\, \right]\frac{\dot{I}}{I}\,,
\label{balancing_linear_hesitancy}
\end{equation}
which generalizes (\ref{balancing_linear}).
Vaccine hesitancy thus introduces an explicit
age dependency, given that cohorts of age $a$
are defined via $a=100-\delta\!A$. The 
basic message of the balancing without vaccine 
hesitancy thus remains, namely that it takes a 
vaccination campaign that is at least two times 
quicker than the percentage spread of the disease.

\section{Evolution of vaccination campaigns}

It is not possible to vaccinate the entire 
population instantly, because vaccines have 
to be first mass-produced and then distributed.
This is illustrated in Fig.~\ref{fig_dailyVaccinations},
where the daily vaccination rates are shown for 
a range of selected countries. Daily rates may vary,
in particular for smaller countries, when
larger batches are delivered from abroad. 
But for larger entities,
like the EU or the US, the trend is linear. 
Overall, vaccination rates can be expected to
track deliveries, with eventual organizational 
problems leading only to temporary delays.

Over the course of several months, the
daily vaccination rates shown in 
Fig.~\ref{fig_dailyVaccinations} rise
roughly linearly during the initial states
of the campaign, 
in line with steadily increasing production 
capacities.\footnote{This linear ramp-up has
been predicted \cite{gros2021covideconomicsrace}.
The reason is that Covid-19 vaccine were
ordered ahead of their approval in large batches. 
But ramping up production capacities implies 
substantial adjustment costs. Minimizing these adjustment
costs, subject to fulfilling the order within a 
contracted time period, leads to a linear ramping-up of 
production capacities \cite{gros2021incentives}.}
Given these considerations, and the data presented
in Fig.~\ref{fig_dailyVaccinations}, we assume that
the number of daily jabs, viz the vaccination 
rate, increases linearly. The fraction of 
the population $v$ vaccinated top down 
increases then with the square of time $t$,
\begin{equation}
v=v(t) = \frac{1}{2}\left(\frac{t}{t_0}\right)^2
 = \frac{p}{2}(\delta\!A)^2,
\qquad\quad
\delta\!A =\frac{t}{t_0\sqrt{p}}\equiv a_0g_v t\,.
\label{v_t_t1}
\end{equation}
The parameter $t_0$ denotes the time 
needed to vaccinate one half of the population.
Given a linear increase of cohort sizes with age, one finds 
that the age of the youngest cohort that can be 
fully vaccinated, denoted $\delta\!A$, 
falls linearly over time, see (\ref{rho_x}). 
The factor $a_0$ in the last definition is 
the characteristic age determining the 
exponential functionality of the IFR, as 
defined by (\ref{IFR}). 

An order of magnitude estimate for the length 
of vaccination campaigns, $t_0$, can be evaluated
from available data. For example, in Israel 
it took about 10 weeks (from the beginning of 
January of 2021 to mid-March 2021) to fully vaccinate 
half the population, resulting in an estimate 
of $t_0 = 10$ (weeks). In the EU, only about 5 
percent of the population was fully vaccinated 
over the same period corresponding to an estimate of
$t_0 =10\sqrt{10}\approx32$ (weeks).\footnote{The 
time needed to vaccinate the entire population, i.e.\ 
to the point $v=1$, is equal $\sqrt{2} t_0$. The parameter 
$t_0$ thus does not denote the full length of the
vaccination campaign, but the time needed to vaccinate
50\% of the population. At that point more than 99\% 
of the fatalities can be avoided and NPIs can be lifted. 
$t_0$ provides thus a good parametrization of the 
effective length of a vaccination campaign.}

The balancing condition (\ref{balancing_linear})
for the case of linear population density
can be used to write the sweet spot at which 
medical costs are kept from rising, in terms 
of the two structural parameters $a_0$ and $p$,
\begin{equation}
\frac{t}{t_0} = 
\left(a_0^2p+\frac{t}{t_0}a_0\sqrt{p}\,\right)t_0\frac{\dot{I}}{I},
\qquad\quad
v_0 = \frac{a_0^2p}{2}\,,
\label{delta_C_p_a0}
\end{equation}
see (\ref{IFR}) and (\ref{rho_x}).\footnote{We 
concentrate on the general analytical solution in 
order to avoid having to make too many specific 
assumptions about the way the pandemic spreads. 
For a more detailed, structural approach 
see \cite{shen2021projected}.} 

As above, one could substitute the parameter $a_0$ with
$a_{0h}$ if the aim is to keep hospitalisations under 
control. Given that equation (\ref{delta_C_p_a0}) 
contains the square of the parameter it follows that it 
will take about 4 times as long to reach the sweet spot 
in terms of hospitalizations than in terms of fatalities.

\section{Discussion}

A key aim for policy makers grappling with a 
continuing outbreak, even when an increasing proportion 
of the population is being vaccinated, is to 
`flatten the curve', i.e.\ 
to keep hospitalizations and fatalities from 
rising exponentially \cite{moore2021vaccination}.
We identified a balancing condition, or 'sweet spot',
at which the health costs remain constant as the 
most vulnerable are being vaccinated first. 
The balancing condition can be interpreted as indicating 
the path of relaxing NPIs along which infections can 
still increase, but fatalities or death remain under control.

Our aim is not to provide detailed
epidemiological modeling and simulations.
Instead, we have shown that three key factors can 
be combined into a simple formula that 
determines the impact of vaccination campaigns
with regard to the time evolution of medical 
costs. First, the mortality risk from a Covid-19 
infection (and that of hospitalisation) increases 
exponentially with age. Second, the sizes 
of age cohorts decrease from the top of the 
population pyramid. 
Third, vaccination proceeds at an increasing speed.

We find considerable differences across 
two dimensions: the form of the age pyramid 
and the measure of health costs: hospitalisations 
or fatalities.  
\begin{itemize}
\item
Age pyramid: Older countries need to vaccinate 
more quickly than younger ones. The difference 
between a typical European country and a high 
fertility country in Sub-Saharan Africa can be 
a factor of 6.
\item
Deaths or hospitalisations: For an old country, 
aiming at keeping hospitalisations constant entails 
a requirement of a vaccination rate about 4 times 
larger than if the aim is to keep fatalities 
constant. For a young country the difference 
is even larger.  
\end{itemize}
Our discussion has focused on the case of Europe, but the formula we derive holds generally and can take into account the wide differences one observes in the speed of vaccination campaigns and the age structure of the population. Moreover, our approach is general enough to accommodate the emergence of different strains of the virus which might increase contagion and/or the risk of severe courses.  For example, the impact of the diffusion of the so-called delta variant, whose diffusion has recently been modelled \cite{gilmour2021preliminary}, would show up in a higher rate of growth infections, which would need a correspondingly higher speed of vaccination to offset its higher contagiousness.

\subsection{Limitations}

In our framework, `vaccinated' implies full immunity,
which is attained for most Covid-19 vaccines only
after the second jab. A reduced levels of immunity, 
like 95\%, is equivalent to an equivalent degree of
vaccine hesitancy, which we also discuss.
Note that only a proportion of the jabs, and not all,
is administered following a strict age criterion 
\cite{cook2021impact}.
Vaccine hesitancy and other factors, such as 
waning immunity with age, can reduce the overall 
effect of vaccination campaigns. Moreover, there 
are other, less age specific costs of the disease, 
like `long Covid' \cite{sudre2021attributes}.
Incorporating these factors would refine the 
model, at the same time making it necessary 
to estimate an increased number of parameters.
See in this regard, e.g., 
\cite{RKIdritteWelle} and \cite{bosetti2021race}
for recent contributions using state-of-the-art
epidemiological models.

We have concentrated on the fact that vaccination 
basically eliminates the risk of death \cite{mateo2021risk}.
Vaccination reduces however also the spread of 
the virus \cite{rossman2021patterns}.
This provides an additional element which 
increases the importance of vaccination speed. 
However, this element becomes significant primarily
in later stages of a vaccination campaign, 
beyond the point when the vulnerable groups 
have been vaccinated. 

\section*{Acknowledgements}

This research has been supported by  the European
Union's Horizon 2020 research and
innovation program, PERISCOPE:
Pan European Response to the Impacts
of Covid-19 and future Pandemics and
Epidemics, under grant agreement
No.\ 101016233, H2020-SC1-PHE CORONAVIRUS-2020-2-RTD.

\section*{Data Availability}

The electronic supporting material
contains the data analyzed.

\bibliographystyle{unsrt}


\end{document}